\documentclass[showpacs,titlepage,prd,aps,
preprint,
amsmath,byrevtex,nofootinbib]{revtex4}

\usepackage{graphicx}
\usepackage[dvips]{epsfig}
\newcommand{\bea}{\begin{eqnarray}}
\newcommand{\eea}{\end{eqnarray}}

\newcommand{\deltaunodue}{\mbox{$\Delta m_{21}^2 $}}
\newcommand{\deltaunotre}{\mbox{$\Delta m_{31}^2 $}}

\newcommand{\tetaot}{\mbox{$\theta_{13}$}}

\newcommand{\nova}{\mbox{NO$\nu$A} }

\newcommand{\beq}{\begin{eqnarray}}
\newcommand{\eeq}{\end{eqnarray}}

\begin{document}

\preprint{FERMILAB-PUB-06-270-T}
\preprint{hep-ph/0609011}

\vspace*{2cm}

\title{ NO$\nu$A and T2K: The race for the neutrino mass hierarchy}
\author{
Olga Mena$^1$,
Hiroshi Nunokawa$^2$ and
Stephen Parke$^1$
}
\address{$^1$
\sl Theoretical Physics Department,
Fermi National Accelerator Laboratory \\
P.O.Box 500, Batavia, IL 60510, USA}

\address{
$^2$
\sl Departamento de F\'{\i}sica,
Pontif\'{\i}cia Universidade Cat\'olica do Rio de Janeiro,\\
C. P. 39071, 22452-9790, Rio de Janeiro, Brazil}
 
\date{\today}

\vglue 1.4cm

\begin{abstract}
The determination of the ordering of the neutrino masses (the hierarchy) is probably a crucial prerequisite to understand the origin of lepton masses and mixings and to establish their relationship to the analogous properties in the quark sector. Here, we follow an alternative strategy to the usual neutrino--antineutrino comparison in long baseline neutrino oscillation experiments: we exploit the combination of the neutrino-only data from the \nova and  the T2K experiments by performing these two off-axis experiments at different distances but at the same $\langle E \rangle/L$, where $\langle E \rangle$ is the mean neutrino energy and $L$ is the baseline. This would require a minor adjustment to the proposed off-axis angle  for one or both of the proposed
experiments.

\end{abstract}
\maketitle
\section{Introduction}
During the last several years the physics of neutrinos has achieved a
remarkable progress. The experiments with solar~\cite{sol,SKsolar,SNO1,SNO2,SNO3,SNOsalt},  atmospheric~\cite{SKatm}, reactor~\cite{KamLAND}, and also
long-baseline accelerator~\cite{K2K} neutrinos (improved with the recent MINOS~\cite{MINOS} results~\cite{MINOSRECENT}) have provided compelling evidence for the existence of neutrino oscillations, implying non zero neutrino
masses. The present data\footnote{We
  restrict ourselves to a three-family neutrino analysis. The
  unconfirmed LSND signal~\cite{LSND} cannot be explained within this
  context and might require additional light sterile neutrinos or more
  exotic explanations. The ongoing
  MiniBooNE experiment~\cite{miniboone} is going to test the
  oscillation explanation of the LSND result.} require two large
($\theta_{12}$ and $\theta_{23}$) and one small ($\theta_{13}$) angles
in the neutrino mixing matrix~\cite{BPont57}, and at least two mass squared differences,
$\Delta m_{ji}^{2} \equiv m_j^2 -m_i^2$ (where $m_{j}$'s are the neutrino
masses), one driving the atmospheric ($\deltaunotre$) and the other one the solar ($\deltaunodue$) neutrino oscillations. The mixing
angles $\theta_{12}$ and $\theta_{23}$ control the solar and the
dominant atmospheric neutrino oscillations, while $\theta_{13}$ is the
angle limited by the data from the CHOOZ and Palo Verde reactor
experiments~\cite{CHOOZ,PaloV}.

The Super-Kamiokande (SK)~\cite{SKatm} and K2K~\cite{K2K} data are well
described in terms of dominant $\nu_{\mu} \rightarrow \nu_{\tau}$
($\bar{\nu}_{\mu} \rightarrow \bar{\nu}_{\tau}$) vacuum
oscillations. The MINOS Collaboration has recently reported the first neutrino oscillation results from $1.27 \times 10^{20}$ protons on target exposure of the MINOS far detector~\cite{MINOSRECENT}. The value of the oscillation parameters from MINOS are consistent with the ones from K2K, as well as from SK data. A recent global fit~\cite{thomas} (see also Ref.~\cite{newfit}) provides the following $3 \sigma$ allowed ranges for the atmospheric mixing parameters
\beq 
\label{eq:range}|\deltaunotre| =(1.9 - 3.2)\times10^{-3}{\rm eV^2},~~~~
0.34<\sin^2\theta_{23}<0.68~.
\eeq
The sign of $\deltaunotre$, sign$(\deltaunotre)$, 
cannot be determined with the existing data. The two possibilities,
$\deltaunotre > 0$ or $\deltaunotre < 0$, correspond to two different
types of neutrino mass ordering: normal hierarchy and inverted hierarchy. Both possibilities are illustrated in Fig.~1, extracted from Ref.~\cite{pom1}. In addition, information on the octant in which $\theta_{23}$ lies, if $\sin^22\theta_{23} \neq 1$, is beyond the reach of present experiments. 

The 2-neutrino oscillation analysis of the solar neutrino data,
including the results from the complete salt phase of the Sudbury
Neutrino Observatory (SNO) experiment~\cite{SNOsalt}, in combination
with the KamLAND spectrum data~\cite{KL766}, shows that the solar neutrino oscillation parameters lie in the low-LMA (Large Mixing Angle) region, with best fit values~\cite{thomas} $\deltaunodue =7.9 \times 10^{-5}~{\rm eV^2}$ and $\sin^2 \theta_{12} =0.30$.


A combined 3-neutrino oscillation analysis of the solar, atmospheric,
reactor and long-baseline neutrino data~\cite{thomas} constrains the third mixing angle to be $\sin^2\theta_{13} < 0.041$ at the $3\sigma$ C.L. However, the bound on $\sin^2 \theta_{13}$ is dependent on the precise value of $\Delta m^2_{31}$.

The future goals for the study of neutrino properties is to
precisely determine the already measured oscillation parameters
and to obtain information on the unknown ones: namely $\theta_{13}$,
the CP--violating phase $\delta$ and the type of neutrino mass
hierarchy (or equivalently sign$(\deltaunotre)$). 

In this paper we concentrate on the extraction of the neutrino mass hierarchy 
by combining the Phase I (neutrino-data only) of the long-baseline $\nu_e$ appearance experiments  T2K~\cite{T2K} and NO$\nu$A~\cite{newNOvA}, both exploiting the off-axis technique\footnote{A neutrino beam with narrow energy spectrum can be produced by placing the detector off-axis, i.~e., at some angle with respect to the forward direction. The resulting neutrino spectrum is very narrow in energy (nearly monochromatic, $\Delta E /E \sim 15 - 25\%$) and peaked at lower energies with respect to the on-axis one. The off-axis technique allows a discrimination between the peaked $\nu_e$ oscillation signal and the intrinsic $\nu_e$ background which has a broad energy spectrum~\cite{adamoff}.}. We start by presenting the general formalism in Sec.~\ref{formalism}. In Sec.~\ref{strategy} we describe the strategy and the experimental setups. In Sec.~\ref{optimal} the experimental configuration is optimized in order to maximize the sensitivity to the mass hierarchy extraction and we present our results for different possible experimental combinations. Finally, in Sec.~\ref{conclusions}, we make our final remarks.
For our analysis, unless otherwise stated, we will use a
representative value of $|\deltaunotre| = 2.4 \times 10^{-3} \
\rm{eV}^2$ and $\sin^2 2 \theta_{23}=1$. 
For the solar oscillation parameters $\deltaunodue$ and $\theta_{12}$, we will use the best fit values
quoted earlier in this section. 

\begin{figure}[t]
\begin{center}
\includegraphics[width=6in]{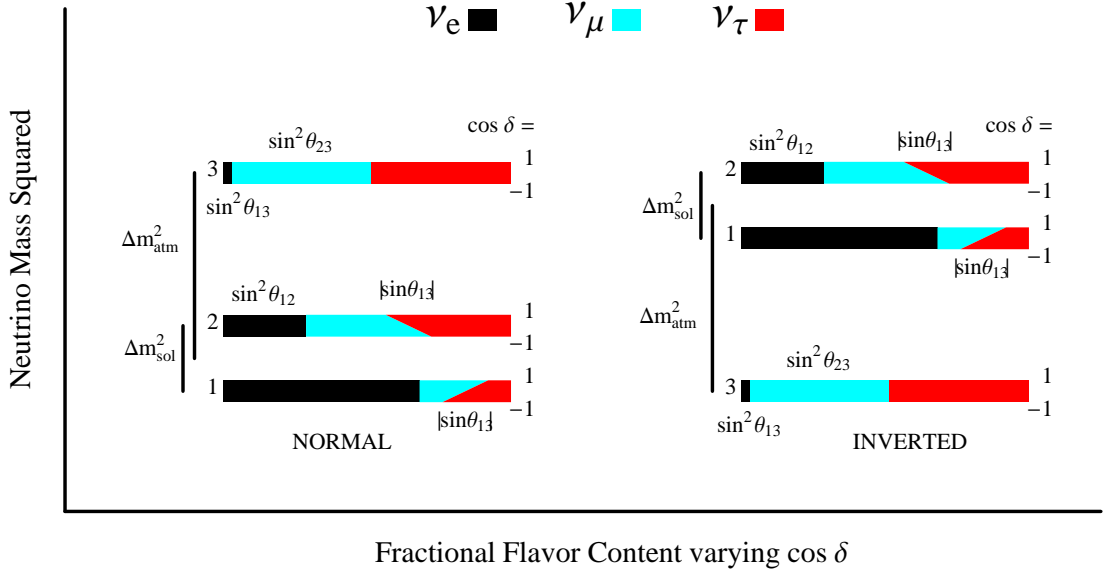}
\caption[]{\it The range of probability of finding the $\alpha$-flavor in the i-th mass eigenstate as indicated for the two different mass hierarchies for the best fit values of the solar and atmospheric mixing parameters, $\sin^2 2 \theta_{13}=0.1$ as the CP-violating phase, $\delta$ is varied.}
\end{center}
\label{fig:hierarchy2}
\end{figure}

\section{The oscillation probability {\boldmath $\nu_\mu \rightarrow \nu$}$_\mathbf{e}$} 
\label{formalism}
The mixing angle $\theta_{13}$ controls $\nu_\mu \rightarrow
\nu_e$ and $\bar{\nu}_\mu \rightarrow \bar{\nu}_e$ conversions in
long-baseline $\nu_e$ appearance experiments and the $\bar{\nu}_e$ disappearance in short-baseline reactor experiments.  Present and future conventional
beams~\cite{MINOS}, super-beams with an upgraded proton source and a
more intense neutrino flux and future reactor 
neutrino experiments have the possibility to
measure\footnote{Smaller values of
this mixing angle could be accessed by very long baseline experiments
such as neutrino factories~\cite{nufact,CDGGCHMR00}, or by
$\beta$-beams~\cite{zucchelli,mauro,BCCGH05,betabeam} which exploit
neutrinos from boosted-ion decays.}, or set a stronger limit on, $\theta_{13}$.
 Therefore, with the possibility of the first measurement of $\theta_{13}$ being made by a 1-to 2-km baseline reactor experiment~\cite{futurereactors}, the long-baseline off-axis $\nu_e$ appearance experiments, T2K~\cite{T2K} and NO$\nu$A~\cite{newNOvA},  need to adjust their focus to emphasize other physics topics.  
The most important of these questions is the form of the mass hierarchy, normal ($\Delta m^2_{31} > 0$)
versus inverted ($\Delta m^2_{31}<0$), and whether or not leptonic CP violation occurs.
 
Consider the probability $P (\nu_\mu \to \nu_e) $ in the context of three-neutrino mixing in the presence of matter~\cite{matter}, represented by the matter parameter $a$, defined as $a \equiv G_{F} n_e/\sqrt{2}$, where $n_e$ is the average electron number density over the baseline, taken to be constant throughout the present study. Defining $\Delta_{ij} \equiv \frac{\Delta m^2_{ij} L}{4 E}$, a convenient and
precise approximation is obtained by expanding to second order 
in the following small parameters: 
$\tetaot$, $\Delta_{21}/\Delta_{32}$, $\Delta_{21}/aL$ and $\Delta_{21}$. 
The result is (details of the calculation can be found in Ref.~\cite{CDGGCHMR00}, see also Ref.~\cite{3probnew}):     
\bea
                  &&P (\nu_ \mu \rightarrow \nu_e) \simeq \nonumber \\
&&\left|\sin \theta_{23}\sin 2 \theta_{13}
\left(\frac{\Delta_{31}}{\Delta_{31}-aL}\right)\sin(\Delta_{31}-aL)e^{-i(\Delta_{32}+\delta)}+\cos\theta_{23}\sin 2 \theta_{12}\left(\frac{\Delta_{21}}{aL}\right) \sin \left( aL \right) \right|^2 \nonumber\\
 &=& 
\sin^2 \theta_{23} \, \sin^2 {2 \theta_{13} } \left(
\frac{\Delta_{31}}{\Delta_{31}-aL} \right)^2     
\sin^2 \left(\Delta_{31} - aL\right) \nonumber \\
&+& \sin {2 \theta_{13} } \sin {2 \theta_{23}} 
\sin 2 \theta_{12} \left(\frac{\Delta_{21}}{aL}\right) \left(\frac{\Delta_{31}}{\Delta_{31}-aL}\right)
\sin \left( a L \right) \sin \left(\Delta_{31}-aL \right) 
 \cos \left(\Delta_{32} + \delta \right) \nonumber \\
 &+& \cos^2 \theta_{23} \sin^2 {2 \theta_{12}} \left(
 \frac{\Delta_{21}}{aL} \right)^2 \sin^2 \left( a L \right), 
\label{eq:probappr}
\eea
where $L$ is the baseline and $a\to -a$, $\delta\to -\delta$ for $P (\bar{\nu}_ \mu \rightarrow \bar{\nu}_e)$.  

Suppose $P(\nu_\mu \to \nu_e) < P(\bar{\nu}_\mu \to \bar{\nu}_e)$ for given energy and baseline, then in vacuum this implies CP violation. However, in matter, this implies CP violation only for the normal hierarchy but not necessarily for the inverted hierarchy around the first oscillation maximum. The different index of refraction for neutrinos and antineutrinos induces differences in the $\nu$, $\bar{\nu}$ propagation that could be misinterpreted as CP violation~\cite{matterosc}. 
Typically, the proposed long baseline neutrino oscillation experiments have a single far detector and plan to run with the beam in two different modes, neutrinos and antineutrinos. In principle, by comparing the probability of neutrino and antineutrino
flavor conversion, the values of the CP--violating phase $\delta$ and
of sign$(\deltaunotre)$ could be extracted. However, different sets of values
of CP--conserving and violating parameters, ($\theta_{13}$, $\theta_{23}$,
$\delta$, sign$(\deltaunotre)$), lead to the same probabilities of
neutrino and antineutrino conversion and provide a good description of
the data at the same confidence level. This problem is known as the
problem of degeneracies in the neutrino parameter
space~\cite{FL96,BCGGCHM01,MN01,BMWdeg,deg} and severely affects the
sensitivities to these parameters in future long-baseline experiments.
Many strategies have been advocated to resolve this issue. Some of the
degeneracies might be eliminated with sufficient energy or baseline
spectral information~\cite{CDGGCHMR00}. In practice, statistical errors
and realistic efficiencies and backgrounds limit considerably the
capabilities of this method. Another
detector~\cite{MN97,BCGGCHM01,silver,BMW02off,twodetect} or
the combination with another
experiment~\cite{BMW02,HLW02,MNP03,otherexp,mp2,HMS05,M05,huber1,huber2} would,
thus, be necessary\footnote{New approaches which exploit other
  neutrino oscillations channels such as muon neutrino disappearance
  have been proposed~\cite{hieratm} for determining the type of
  hierarchy. They require very precise measurements of the neutrino oscillation disappearance oscillation parameters though. }.

The use of only a neutrino beam could help in resolving the type of
hierarchy when two different long-baselines are 
considered~\cite{HLW02,MNP03,SN1,twodetect,SN2}. 
It was shown in ref.~\cite{MNP03} that if the $\langle E\rangle/L$ for the two
different experiments is approximately the same then the allowed regions
for the two hierarchies are disconnected and thus this method for determining
the hierarchy is free of degeneracies.\footnote{This method, while optimal 
 for extracting the neutrino mass hierarchy, is not the most powerful one for extracting the CP violating phase $\delta$ for which antineutrino running would be, in general,  necessary.}  Naively, we can understand this 
method in the following way for $\sin^2 2 \theta_{13}>0.01$: 
assume that matter effects are negligible for the short baseline, 
then at the same $\langle E\rangle/L$, if the oscillation probability 
at the long baseline is larger than the oscillation probability 
at the short baseline, one can conclude that the hierarchy is normal, since 
matter effects enhance the neutrino oscillation probabilities for 
the normal hierarchy.
For the inverted hierarchy the oscillation probability 
for the long baseline
is suppressed relative to the short baseline. This will be explained in some detail in the
next section.

\section{ our strategy: only neutrino running and two detectors}
\label{strategy}

Following the line of thought developed by Minakata, Nunokawa and Parke~\cite{MNP03}, we exploit the neutrino data from two experiments at different distances and at different off-axis locations. The off-axis location of the detectors and the baseline must be chosen such that the $\langle E \rangle /L$ is the same for the two experiments. Here we explain the advantages of such a strategy versus the commonly exploited neutrino-antineutrino comparison.

\begin{figure}[t]
\begin{center}
\begin{tabular}{ll}
\includegraphics[width=3in]{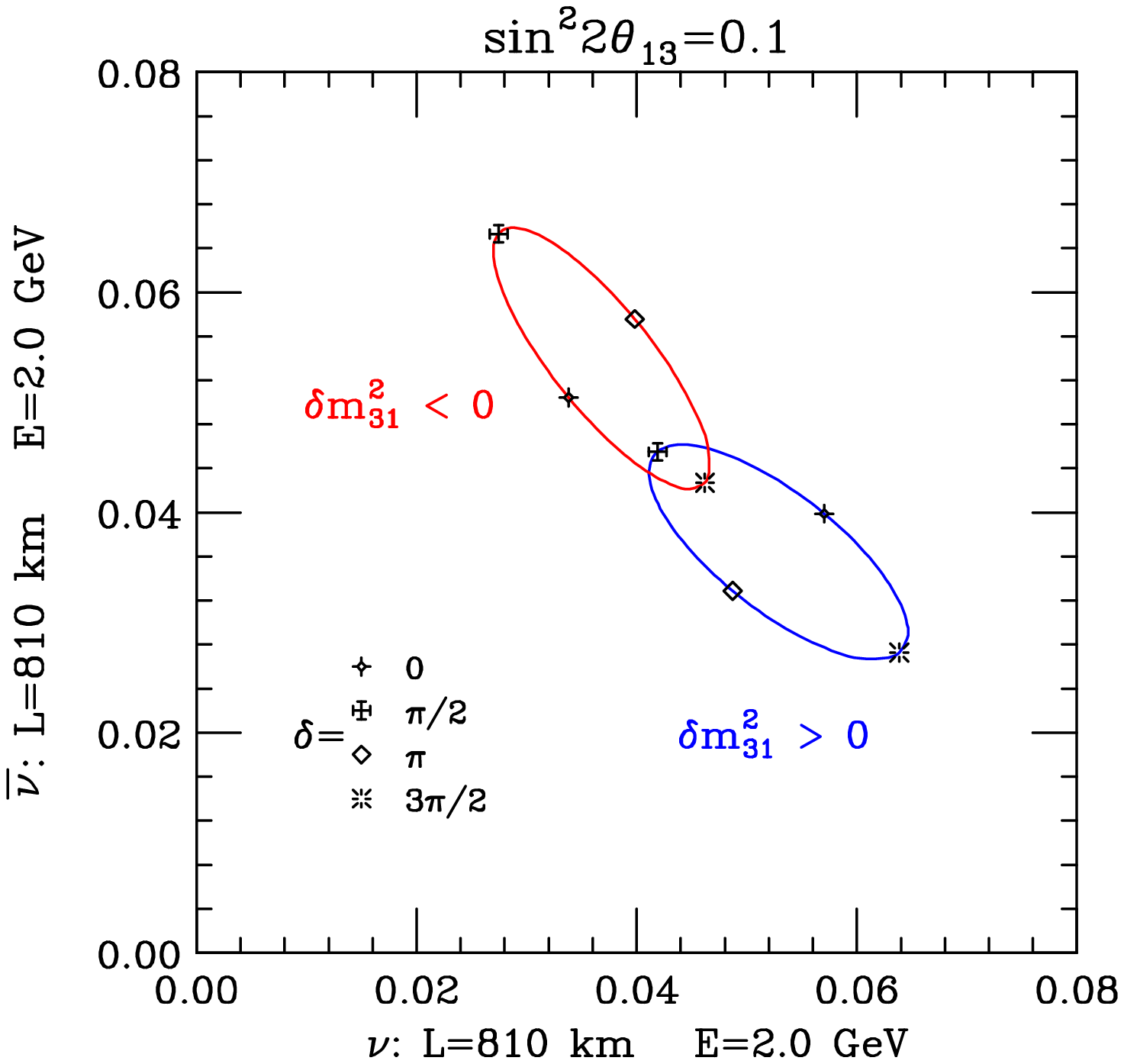}&\hskip 0.cm
\includegraphics[width=3in]{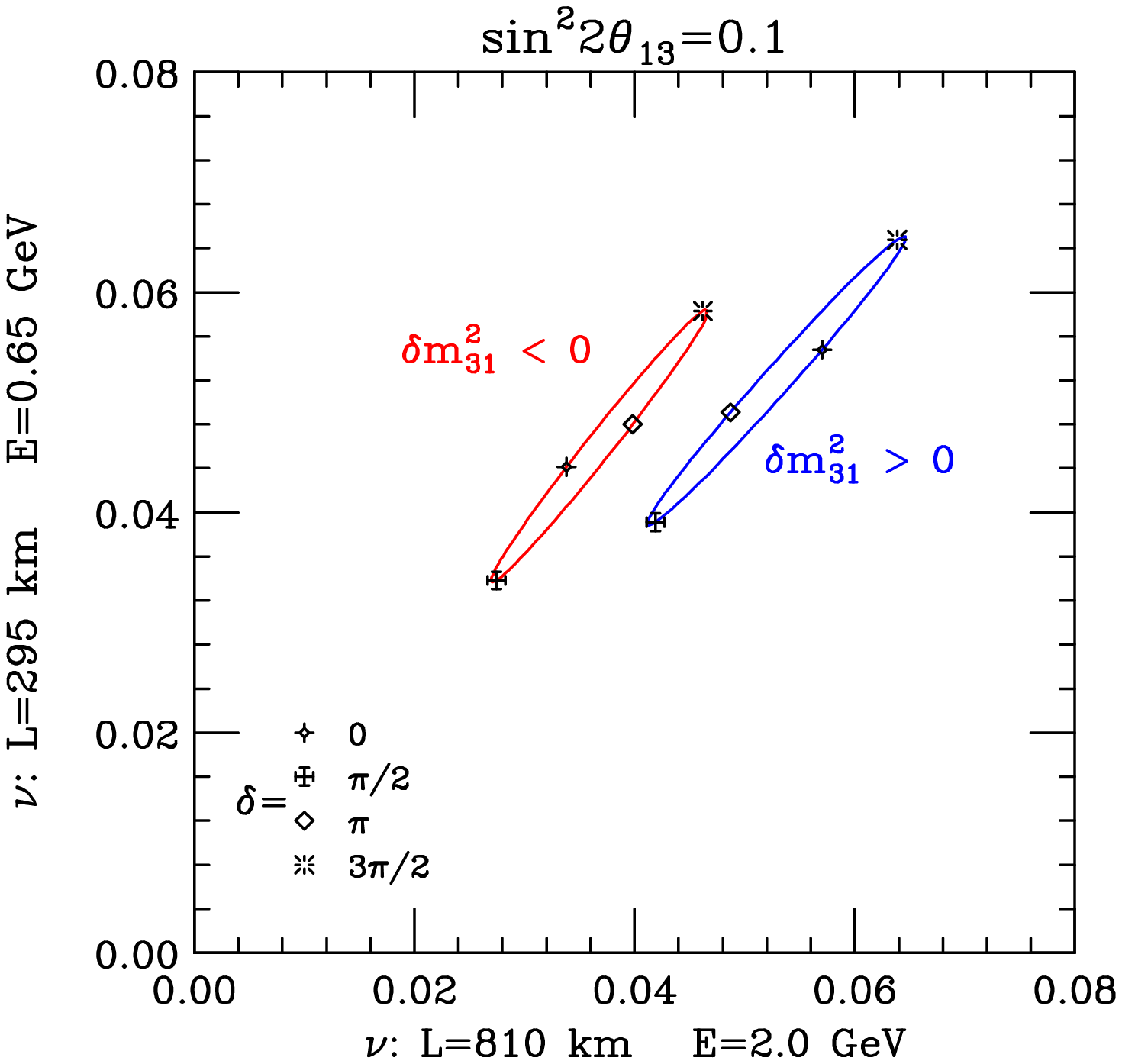}\\
\hskip 2.truecm
{\small (a) Neutrino--Antineutrino}            &
\hskip 2.truecm
{\small (b) Neutrino--Neutrino}
\end{tabular}
\end{center}
\caption[]{\textit{(a) Bi--probability neutrino--antineutrino ellipses at the far distances (810 km) for normal (lower blue) and inverted (upper red)
hierarchies.  (b) Same as (a) but in the bi--neutrino plane at the short (295 km) and far distances (810 km).}}
\label{fig:comp}
\end{figure}

 Suppose we compute the oscillation probabilities $P(\nu_ \mu \to \nu_e)$ and $P(\bar \nu_\mu \to \bar \nu_e)$ for a given set of oscillation parameters and the CP-phase $\delta$ is varied between $0$ and $2 \pi$: we obtain a closed CP trajectory (an ellipse) in the bi--probability space of neutrino and antineutrino conversion~\cite{MN01}. Matter effects are responsible for the departure of the center of the ellipses from the diagonal line in the bi--probability plane for the both cases of normal and inverted hierarchy, see Fig.~\ref{fig:comp}~(a), where we have illustrated the case for $E=2.0$ GeV and $L=810$ km. The distance between the center of the ellipse for the normal hierarchy (lower blue) and that for  the inverted hierarchy (upper red) is governed by the size of the matter effects. Notice that the ellipses overlap for a significant fraction of values of the CP--phase $\delta$ for every allowed value of $\sin^2 2
\theta_{13}$.  This indicates that, generically, a measurement of the probability of conversion for neutrinos and antineutrinos cannot uniquely determine
the type of hierarchy in a single experiment. This makes the determination of sign$(\deltaunotre)$ extremely difficult, i.~e., the sign$(\deltaunotre)$-extraction is not free of degeneracies.

In the case of bi--probability plots for neutrino--neutrino modes 
at different distances (which will be referred as near (N) and far (F)), 
the CP--trajectory is also elliptical (see Fig.~\ref{fig:comp}~(b)). 
The overlap of the two bands, which implies the presence of a 
degeneracy of the type of hierarchy with other
parameters, is controlled by the difference in the slopes and the width of the
bands. Using the fact that matter effects are small 
($aL\ll\Delta_{31}$), we can perform a
perturbative expansion and assuming that the $\langle E \rangle /L$ of the near and far experiments is the same\footnote{The reason for this choice of $\langle E \rangle /L$ is explained in the next paragraph.}, at first order, 
the ratio of the slopes reads~\cite{MNP03}
\beq
\frac{\alpha_+}{\alpha_-} \simeq
1 +  4 \left( a_{\rm N} L_{\rm N} - a_{\rm F} L_{\rm F} \right)\left( \frac{1}{\Delta_{31}} - \frac{1}{\tan(\Delta_{31})} \right)~,
\label{eq:ratioapp}
\eeq
where $\alpha_+$ and $\alpha_-$ are the slopes of the center of the ellipses as one varies $\theta_{13}$ for normal and inverted
hierarchies, and $a_{\rm F}$ and $a_{\rm N}$ are the matter parameters, $L_{\rm F}$ and $L_{\rm N}$ are the baselines for the two experiments\footnote{Notice that although we are using the constant
density approximation, $a_{\rm F}$ and $a_{\rm N}$ are 
different because the average density depends on the baseline. 
For T2K (\nova) we use an average density times electron fraction 
equal to $1.15$ ($1.40$) {\rm g $\cdot$ cm}$^{-3}$.}. 
The separation between the center of the ellipses for the two hierarchies increases 
as the difference in the matter parameter times the path length 
for the two experiments increases. 
Also, since $(\Delta^{-1} - \cot{\Delta})$ is a
monotonically increasing function, we conclude that the smaller the energy, the larger the ratio of slopes, at least for the same $\langle E \rangle /L$.

However the width of the ellipses is crucial: even when the separation 
between the central axes of the two regions is substantial if the ellipses for
the normal and inverted hierarchy overlap the hierarchy cannot be resolved for
values of CP phase, $\delta$, for which there is overlap.
The width of the ellipses is controlled by the difference in the 
$\langle E \rangle /L$ of the two experiments.
For fixed $\theta_{13}$ the ellipses are flat if there exists 
two different $\delta$ which give the same oscillation
probability for both the near, $P_{N}$, and far, $P_{F}$, detectors. That is, if
\beq
P_N(\theta_{13}, \delta)=P_N(\theta_{13}, \delta^\prime) \quad &{\rm and} & \quad 
P_F(\theta_{13}, \delta)=P_F(\theta_{13}, \delta^\prime) 
\label{eq:condition}
\eeq
has non-trivial solutions for $\delta$ and $\delta^\prime$.
Eqs.~(\ref{eq:condition}) are satisfied when $\cos(\Delta+\delta)=\cos(\Delta+\delta^\prime)$
for both the near and far experiments simultaneously. Both equations can only be satisfied
when $\Delta_N=\Delta_F$, that is the two experiments have the same
$\langle E \rangle /L$.

Thus, we have two conditions to satisfy to optimize the determination 
of the neutrino mass hierarchy: 
\begin{itemize}
\item (a) maximize the difference in the 
factor $(a L)$ for both experiments and  
\item (b) minimize the ellipses 
width by performing the two experiments at the same $\langle E \rangle /L$.
\end{itemize} 

The most promising way to optimize the sensitivity to the hierarchy with relatively near term data is therefore to focus on the neutrino running mode and to exploit the Phase I data of the long-baseline off-axis $\nu_e$ appearance experiments, T2K  and NO$\nu$A. T2K utilizes a steerable neutrino beam from J-PARC and Super-Kamiokande and maybe eventually Hyper-Kamiokande as the far detector. The beam will peak at $0.65$ GeV with the detector off-axis by an angle of $2.5^\circ$ at $295$ km. For this configuration the matter effects are small but not negligible~\cite{matter_t2k}. 
NO$\nu$A proposes to use the Fermilab NuMI beam with a baseline of 
$810$~km  with a $30$ kton low density tracking calorimeter with an efficiency of $24\%$. Such a detector would be located $12$~km off-axis distance from the beam center at $L=810$ km (corresponding to $0.85^\circ$ off-axis 
angle), resulting in a mean neutrino energy of $2.0$ GeV. Matter effects are quite significant for NO$\nu$A. Therefore, the condition (a) is satisfied, since $(aL)_{\textrm{\nova}} \simeq 3 (aL)_{\textrm{T2K}}$ . What about the condition (b)? A back-of-the-envelope calculation indicates us that the current off-axis angles are not such that $\langle E \rangle /L$ of the two experiments will be the same. However, by placing the detector(s) at slightly different off-axis angle(s),
 one can arrange that the $\langle E \rangle/L$ of the two experiments to be exactly the same. This strategy  would only need half of the time of data taking (because we avoid the antineutrino running), when compared to the standard one (i.e. running in neutrinos and antineutrinos at a fixed energy, $\langle E \rangle$, and baseline, $L$).

\section{optimizing the \nova and T2K detector locations}
\label{optimal}
In this section we present what could be achieved if \nova and T2K setups are carefully chosen, focusing on the physics potential of the combination of their future data. We define the Phase I of the experiments as follows. For the T2K experiment, we consider 5 years of neutrino running and SK as the far detector with a fiducial mass of $22.5$ kton and $70\%$ detection efficiencies. For the \nova experiment, we assume $6.5\times 10^{20}$ protons on target per year, 5 years of neutrino running and the detector described in the previous section.

\begin{figure}[t]
\begin{center}
\begin{tabular}{ll}
\includegraphics[width=3in]{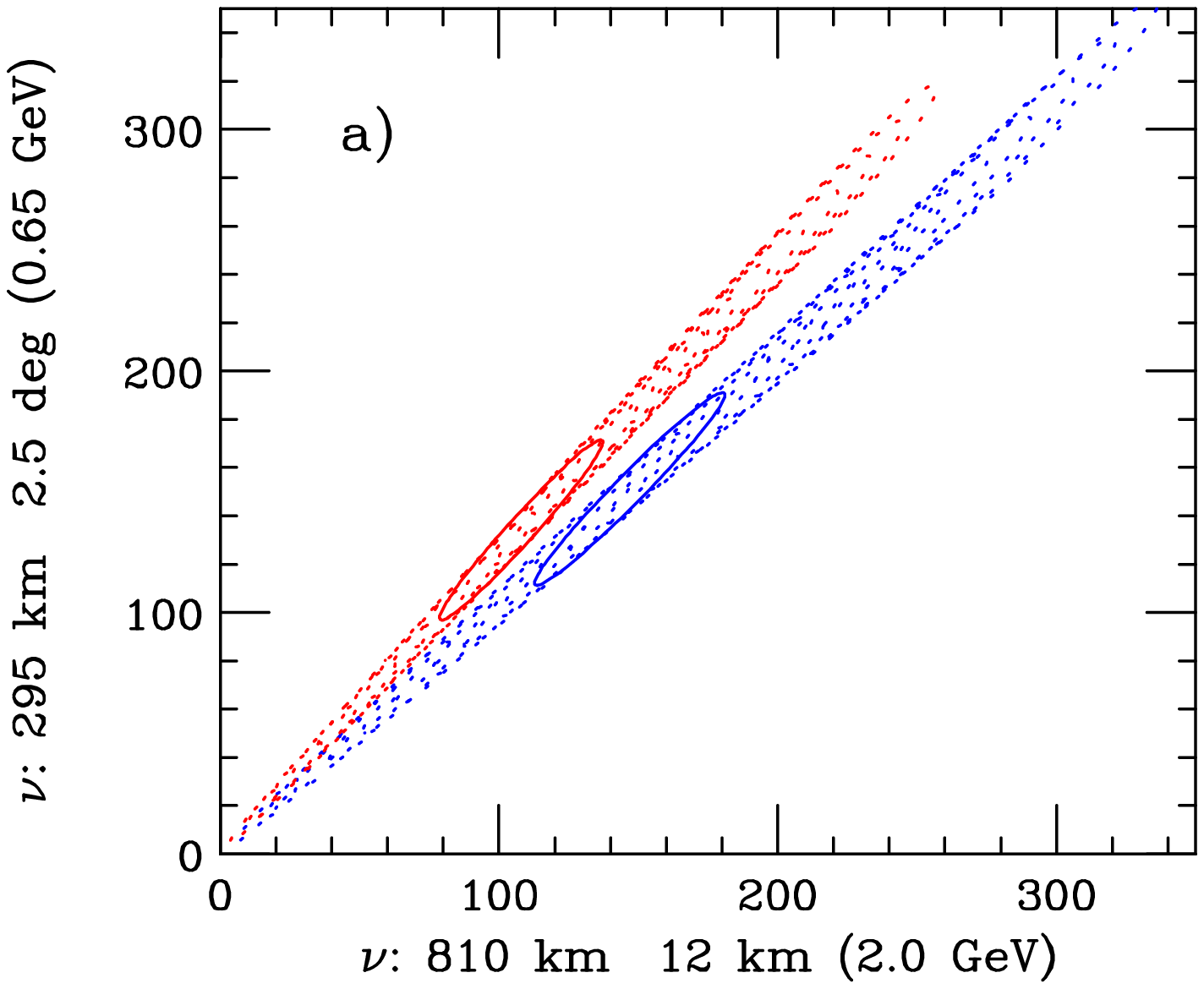}&\hskip 0.cm
\includegraphics[width=3in]{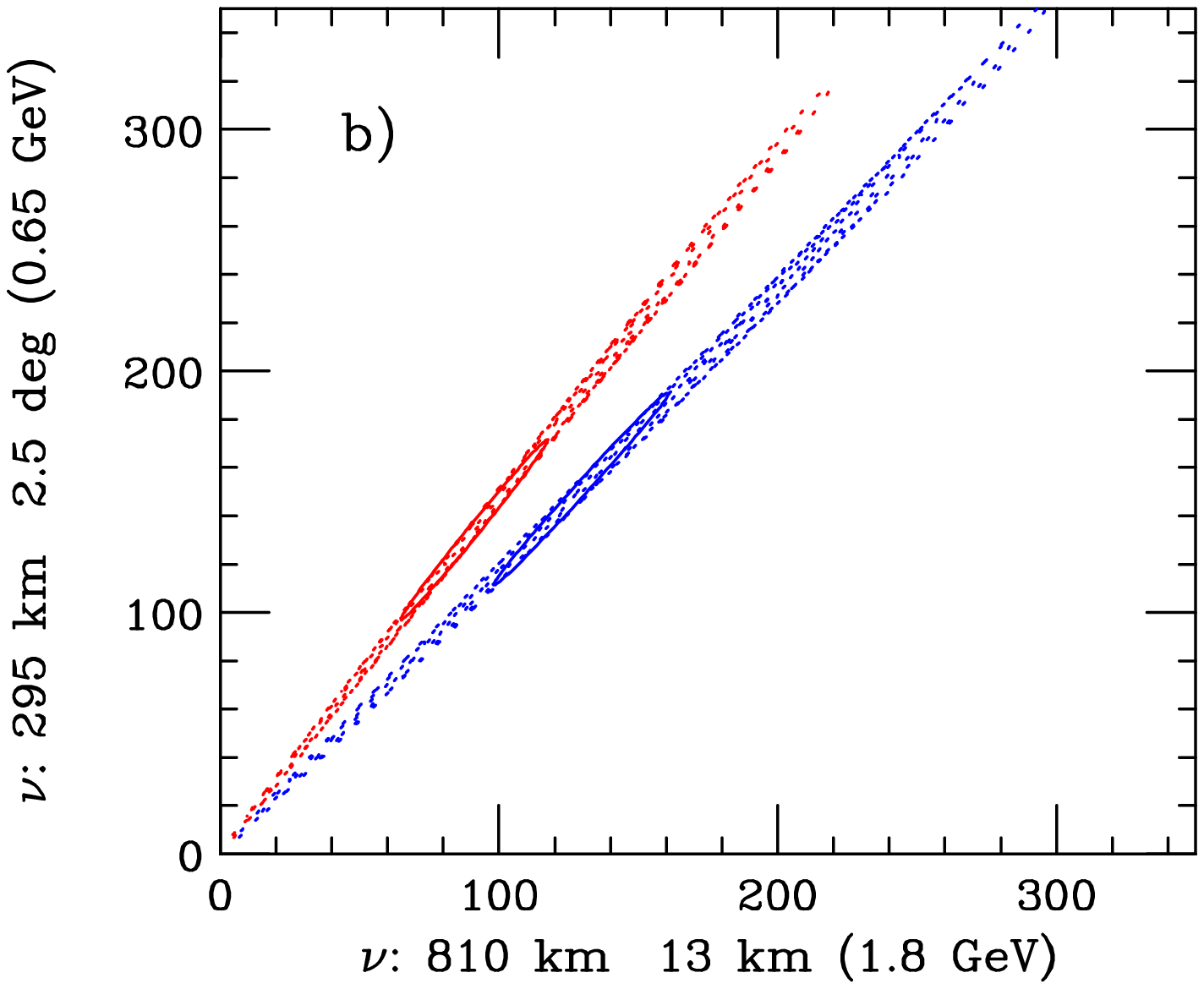}\\
\includegraphics[width=3in]{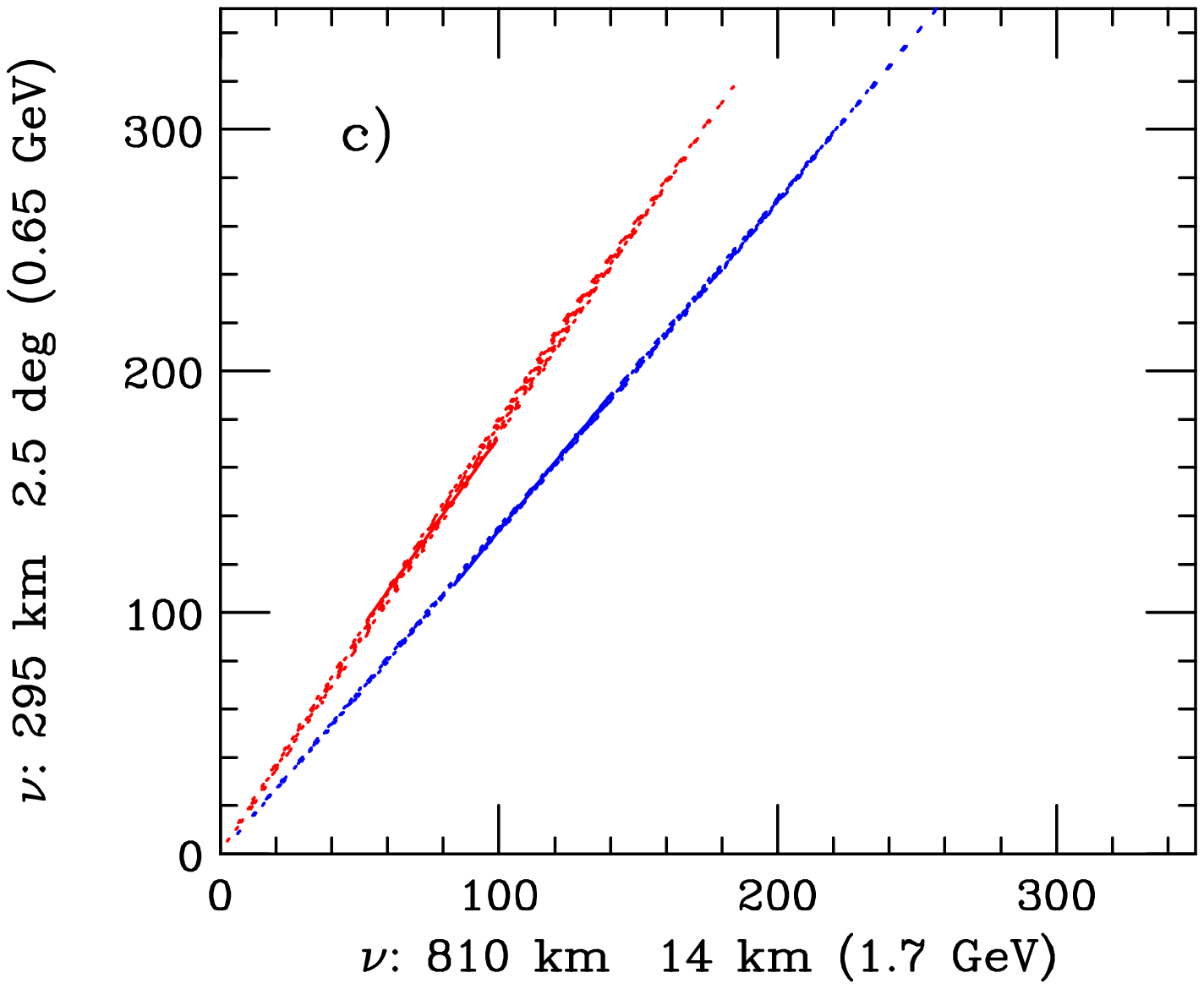}&\hskip 0.cm
\includegraphics[width=3in]{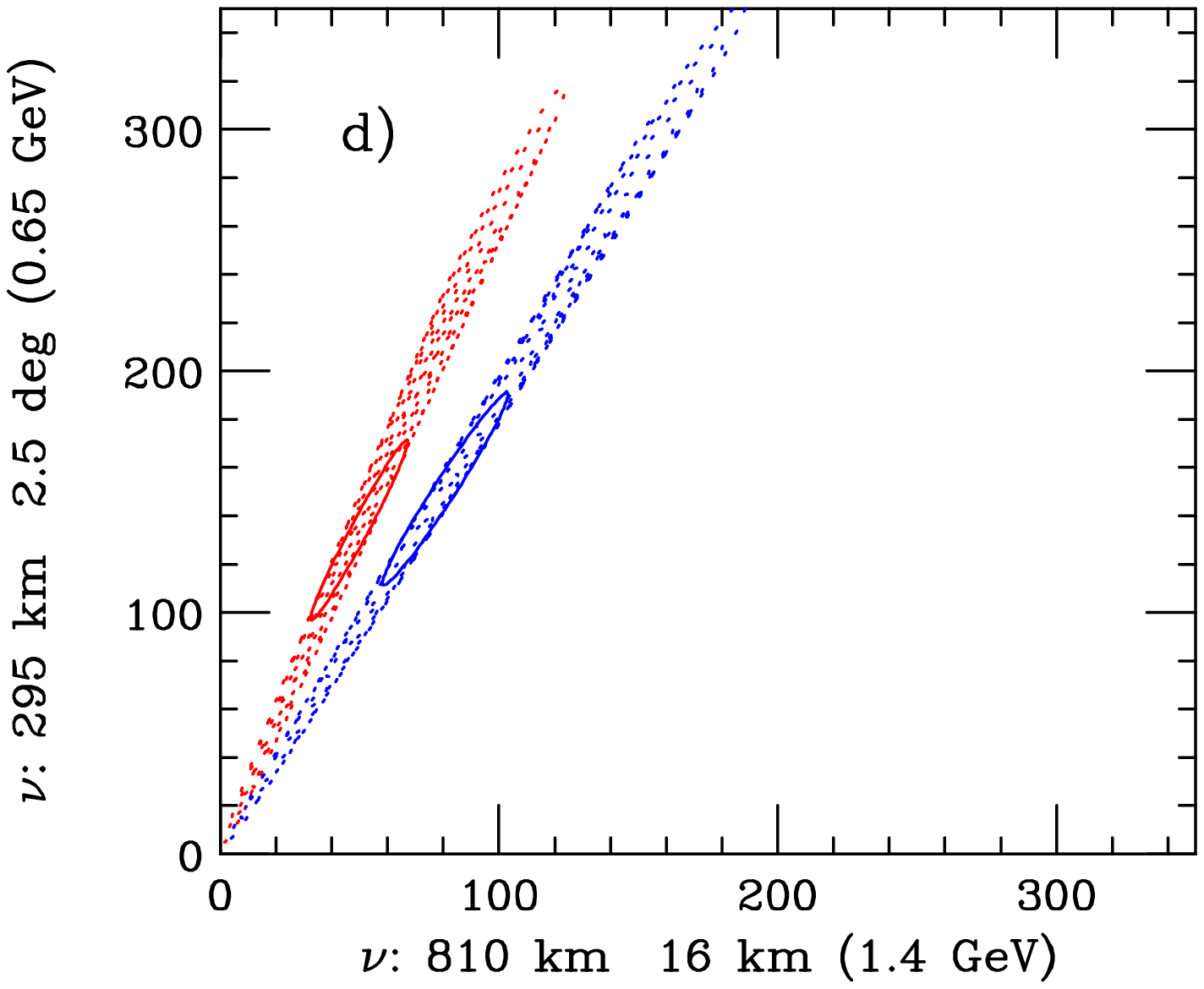}\\
\end{tabular}
\end{center}
\caption[]{\textit{(a)Bi--event neutrino--neutrino ellipses at the \nova and T2K experiments for normal (lower blue) and inverted (upper red)
hierarchies. From bottom up, the dashed ellipses correspond to $\sin^2 2 \theta_{13}$ varying from $0.01$ to $0.2$ with a stepsize of $0.01$ and the solid ellipses illustrate the case
$\sin^2 2 \theta_{13}=0.1$. The T2K far detector is located off-axis by an angle of $2.5^{\circ}$. The \nova detector is placed $12$~km off-axis. (b,c,d) Same as (a) but with the \nova detector located at $13$~km, $14$~km and $16$~km off-axis, respectively.}} 
\label{fig:t2kandnova}
\end{figure}

We present in Figs.~\ref{fig:t2kandnova} the bi--event neutrino--neutrino plots, fixing the off-axis location of the T2K detector to $2.5^\circ$ but varying the off-axis distance of the \nova far detector, where the number of events is computed by integrating the neutrino flux convoluted with the cross section, the oscillation probability and the efficiency over an energy window of $1$ GeV centered around the mean neutrino energy (for both \nova and T2K experiments).
If the location of the \nova far detector is the proposed one, i.e. $12$~km off-axis, the width of the ellipses is not negligible and the determination of the hierarchy would become a really difficult task even if nature has been kind in its choice of $\delta$, see Fig.~\ref{fig:t2kandnova}~(a). As we anticipate in the previous section, to maximize the difference  in the product $aL$, see Eq.~(\ref{eq:ratioapp}), does not ensure a degeneracy-free hierarchy measurement, due to the impact of the ellipses width.
Notice, however, that the width of the ellipses is minimal when the $\langle E \rangle /L$ of both experiments is almost the same: this occurs when the \nova far detector is located $14$ km off axis, see Fig.~\ref{fig:t2kandnova} (c), for which the mean energy $\langle E\rangle \sim 1.7$ GeV. If the \nova far detector is placed $16$~km off-axis, the $\langle E \rangle/L$ of the T2K and \nova experiments is no longer the same and the ellipses width has grown with respect to the $14$~km off-axis setup. \textit{The most important lesson we have learned is that if the T2K beam is off-axis by an angle of $2.5^\circ$, the ideal location for the \nova detector to optimize the hierarchy extraction would be $14$~km off-axis.}  Fig.~\ref{fig:t2kandnova} shows the potential of the combination of the data from the Phase I (only neutrinos) of the T2K and \nova experiments, without relying on future second off-axis detectors (placed at a shorter distance or at the second oscillation maximum) and future upgraded proton luminosities (and/or detection technologies providing almost perfect detection efficiencies). 

The disadvantage is that the configuration at $14$~km off-axis implies a $30\%$ loss in statistics with respect to the \textit{default} configuration at $12$~km off-axis. Consequently, the best strategy would be the following: one could tune the $\langle E \rangle /L$ of the two experiments to be the same by changing the T2K off axis-angle, since the design of the T2K experiment could in principle allow for such a modification, keeping the \nova off-axis distance fixed to $12$~km which corresponds to $\langle E \rangle =2.0$ GeV. In order to reproduce the same $\langle E \rangle/L$ than the corresponding one for the \nova experiment, the energy at T2K should be $0.73$ GeV, which could be easily achieved by placing the detector $2^\circ$ off-axis, for which $\langle E \rangle = 0.75$ GeV.  In principle, since the configuration at $2 ^\circ$ off-axis would also imply a gain in statistics for the T2K experiment with respect to the one at $2.5^\circ$ off-axis, one would expect an improvement in the sensitivity. However, we noticed that the ratio for the two solutions Eq.~(\ref{eq:ratioapp}) increases as the energy decreases. Since the configuration of \nova detector located at $14$~km off-axis and T2K located at $2.5^\circ$ off-axis provides a lower $\langle E  \rangle$ than the energy for the configuration obtained by placing the \nova detector to its default location and moving slightly T2K to $2^\circ$ off-axis, the sensitivity achieved in the former is higher than in the latter.

The data fits we perform here are based on a $\chi^2$ analysis, where the $\chi^2$ function reads:
\begin{equation}
\chi^2=\sum_{i=T2K,NO\nu A}\left(\frac{N^{obs}_{i} - N^{th}_{i}(\sin^2 2\theta_{13}, \delta)}{\sigma_i}\right)^2~,
\end{equation}
where $N^{obs}$ is the measured number of events, $N^{th}(\sin^2 2\theta_{13}, \delta)$ is the expected number of events for a particular choice of the hierarchy and $\sigma$ refers to the statistical error corresponding to $N^{th}$. Our analysis includes the intrinsic  $\nu_e$ background. In our plots we show the $90\%$ confidence level regions in the ($\sin^2 2 \theta_{13}$, $\delta$) plane resulting from the data fits, where the $90\%$ CL is defined as $|\chi_{min}^2$(wrong hierarchy) - $\chi_{min}^2$(true hierarchy)$| > 4.61$ for 2 d.o.f~\cite{pdgstat}. It is very important to notice here that our  $90\%$ CL contours translate into the $97\%$ CL contours for 1 d.o.f.~\footnote{In order to test whether or not the 2 d.o.f approach is more appropriate, we use a Monte Carlo technique in which we simulate $10^6$ experiments for each point in the ($\sin^2 2 \theta_{13}$, $\delta$) parameter space and we compute the probability of distinguishing the hierarchy for each point. The result of the Monte Carlo analysis agrees with the conservative approach of 2 d.o.f but it does not reproduce the 1 d.o.f confidence levels. This Monte Carlo analysis has therefore guided us to use 2 d.o.f.}
For all the sensitivity (exclusion) plots shown in this paper, the systematic error was not included but later we will make some comments on the impact of this error on our results.

\begin{figure}[t]
\begin{center}
\begin{tabular}{ll}
\includegraphics[width=3in]{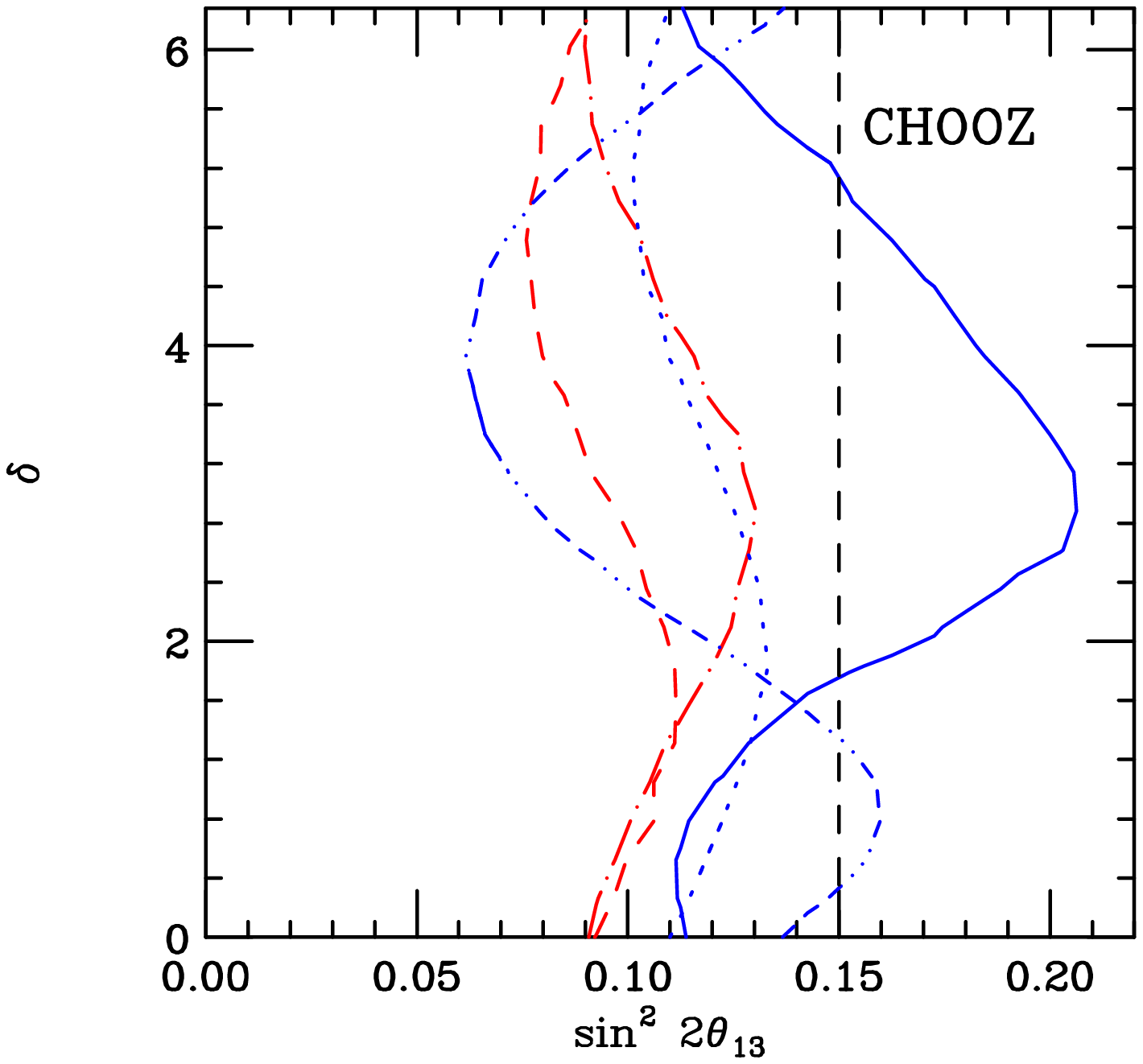}&\hskip 0.cm
\includegraphics[width=3in]{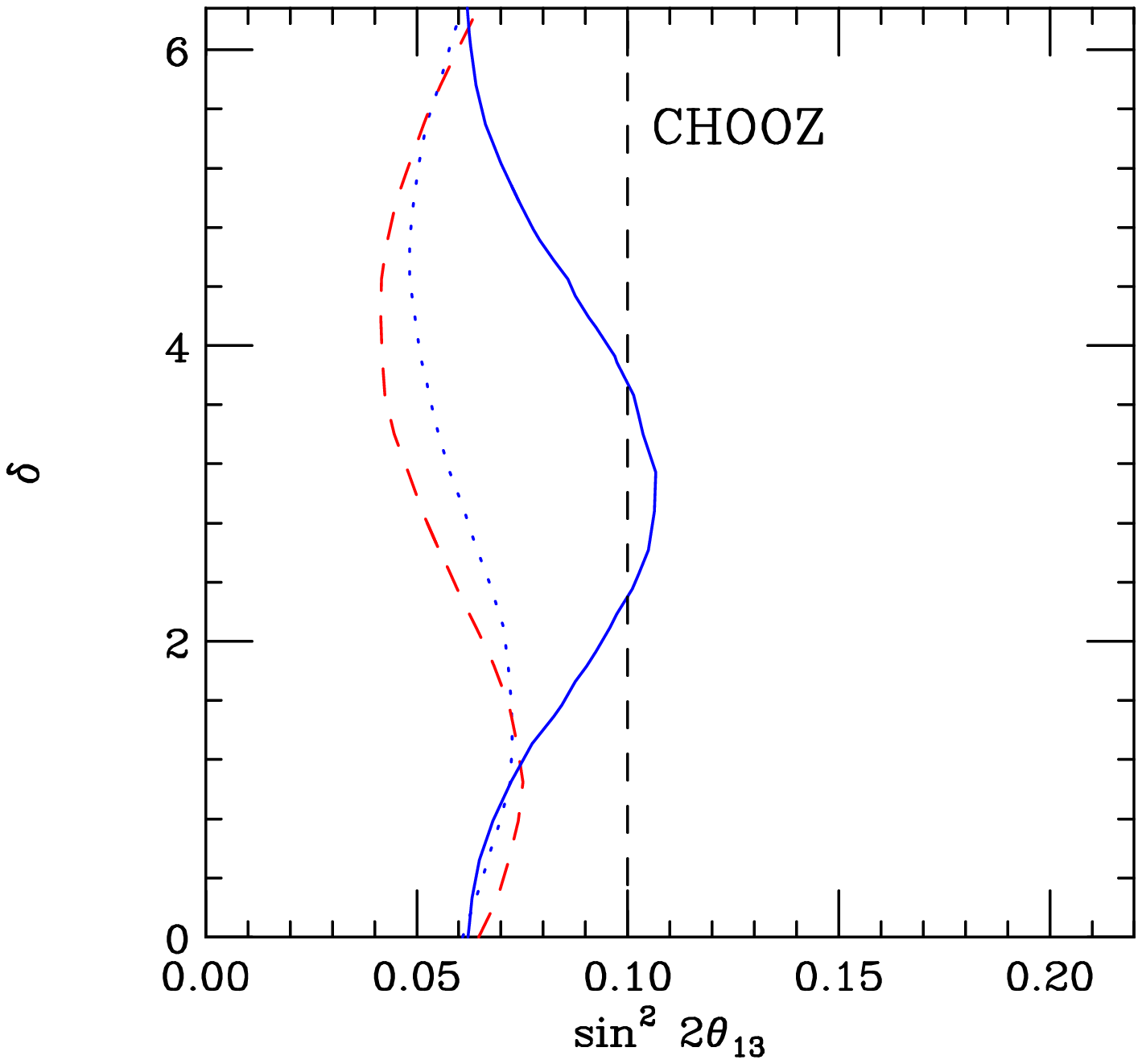}\\
\hskip 2.truecm
{\small (a) $\Delta m^2_{31} = +2.4 \times 10^{-3}$~eV$^2$}            &
\hskip 2.truecm
{\small (b) $\Delta m^2_{31} = +3.0 \times 10^{-3}$~eV$^2$}       \\  
\end{tabular}
\end{center}
\caption[]{\textit{(a) $90\%$ CL (2 d.o.f) hierarchy resolution for different possible combinations: the \textit{default} one (T2K at an off-axis angle of $2.5^\circ$ and \nova far detector at $12$~km off-axis, in solid blue),  T2K at an off-axis angle of $2.5^\circ$ and \nova far detector at $13$~km off-axis (long dash-dot red curve),  T2K at an off-axis angle of $2.5^\circ$ and \nova far detector at $14$~km off-axis (short dashed red curve), T2K at an off-axis angle of $2.5^\circ$ and \nova far detector at $16$~km off-axis (three dots-three dashes blue curve) and  T2K at an off-axis angle of $2^\circ$ and \nova far detector at $12$~km off-axis (dotted blue curve). We have considered the statistics corresponding to the Phase I of both experiments. The vertical dashed line indicates the $95\%$ CL CHOOZ bound for the value of $\Delta m^2_{31}$ for the panel. (b) The same as (a) but assuming that $\Delta m^2_{31}= 3.0 \times 10^{-3} \ eV^2$ and only for the three most representative combinations: the \textit{default} one (in solid blue), T2K at an off-axis angle of $2^\circ$ and \nova far detector at $12$~km off-axis (dotted blue curve) and the optimal one, that is, T2K at an off-axis angle of $2.5^\circ$ and \nova far detector at $14$~km off-axis (short dashed red curve). If one reinterprets these limits for 1 d.o.f then they correspond to the $95\%$ CL, approximately.}}
\label{fig:exclusion}
\end{figure}

We summarize the results in Figs.~\ref{fig:exclusion}, where we present the exclusion plots in the ($\sin^2 2\theta_{13}, \delta$) plane for a measurement of the hierarchy at the $90\%$ CL for the several possible combinations, assuming that nature's solution is the normal hierarchy and $\Delta m^2_{31}=2.4\times 10^{-3}$~eV$^2$ (left panel) and $\Delta m^2_{31}=3.0 \times 10^{-3}$~eV$^2$ (right panel) (in the light of the recent MINOS results, we explore here the impact of a larger  $\Delta m^2_{31}$). We show as well the corresponding CHOOZ bound for $\sin^2 2 \theta_{13}$. A larger value of $\Delta m^2_{31}$ implies more statistics and, consequently, a sensitivity improvement: see Fig.~\ref{fig:exclusion} (b), where for the sake of illustration only three representative configurations are shown.

If both T2K and \nova run in their \textit{default} configurations the combination of their future Phase I data (only neutrinos) will not contribute much  to our knowledge of the neutrino hierarchy, see the solid blue line in Figs.~\ref{fig:exclusion}. If we fix the T2K off-axis location to its \textit{default} value of $2.5^\circ$ but we change the location of the \nova detector to $14$~km the improvement is really remarkable, see the dashed red line in  Figs.~\ref{fig:exclusion}: the sensitivity to the mass hierarchy has a milder dependence on the CP-phase $\delta$ once that the $\langle E\rangle /L$ of the two experiments is chosen to be the same. The best sensitivity to the hierarchy extraction is clearly achieved 
when the \nova experiment is at $14$~km off-axis and the T2K off-axis angle is 
the \textit{default} one. If the T2K off-axis angle is slightly modified to $2^\circ$, see the dotted lines in Figs.~\ref{fig:exclusion} it would be possible to reproduce the results from the combination of the data from T2K located at $2.5^\circ$ off-axis and the \nova detector placed at $13$~km off-axis.

\begin{figure}[h]
\begin{center}
\begin{tabular}{ll}
\includegraphics[width=3in]{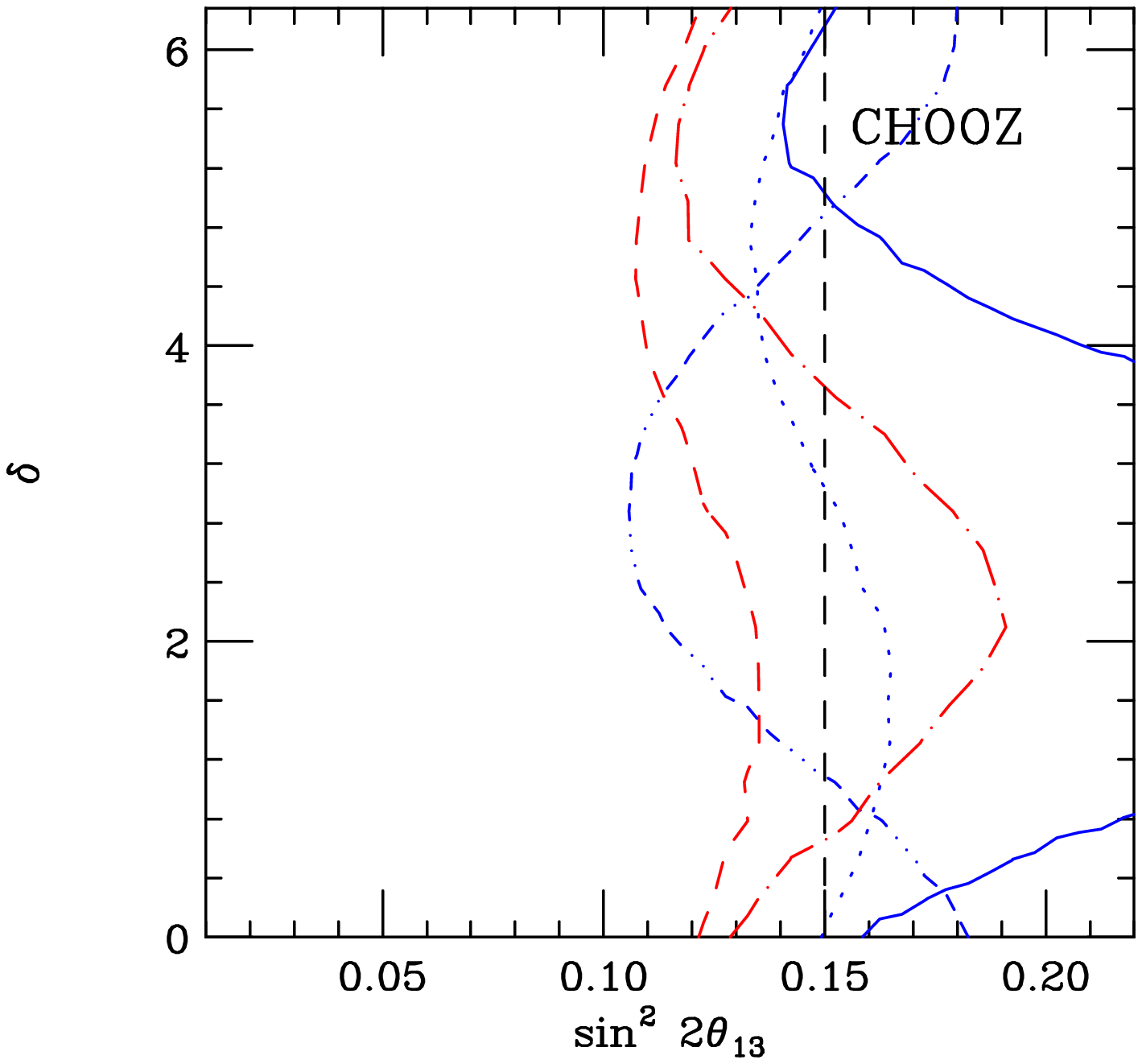}&\hskip 0.cm
\includegraphics[width=3in]{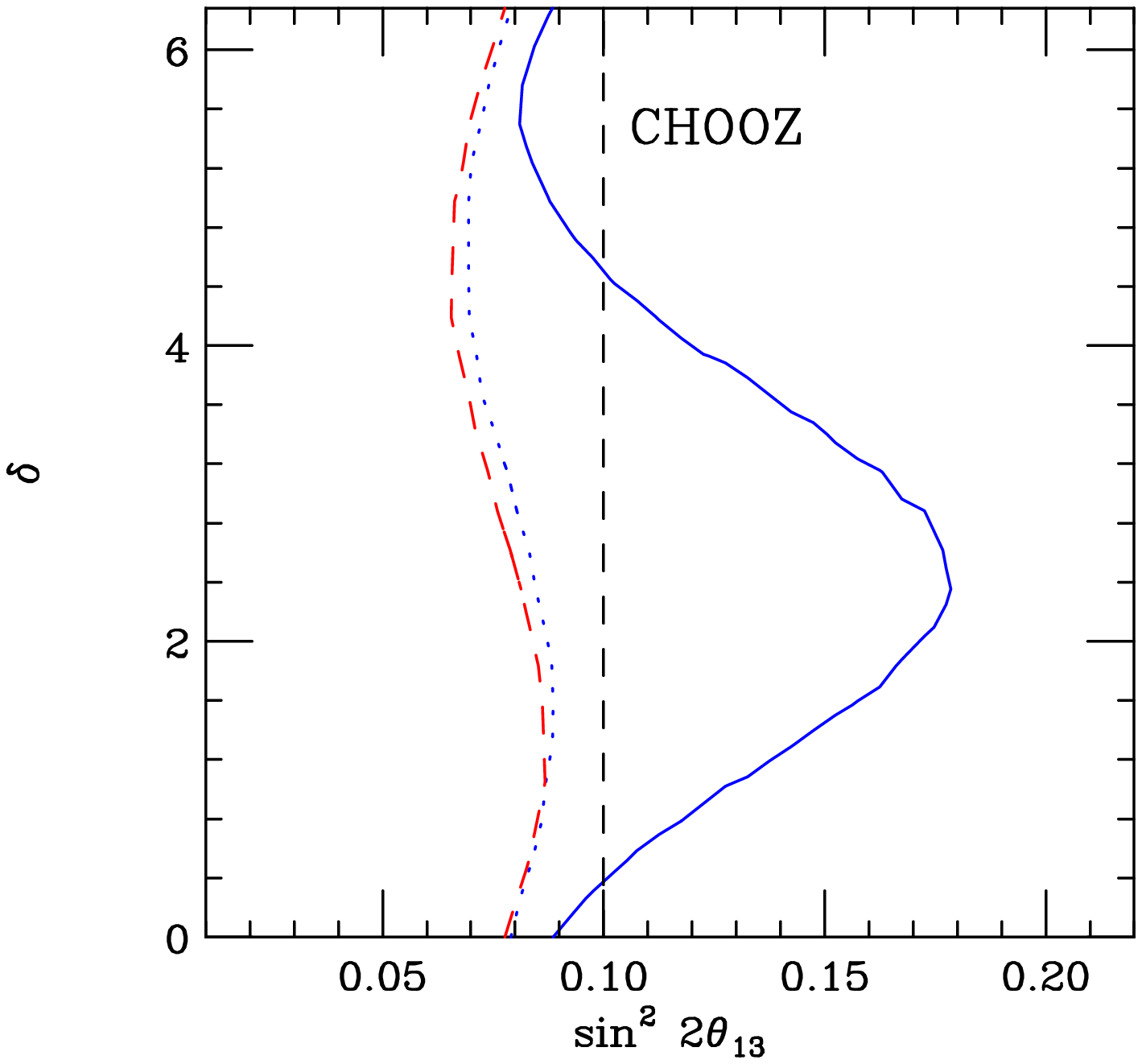}\\
\hskip 2.truecm
{\small (a) $\Delta m^2_{31} = -2.4 \times 10^{-3}$~eV$^2$}            &
\hskip 2.truecm
{\small (b) $\Delta m^2_{31} = -3.0 \times 10^{-3}$~eV$^2$}       \\  
\end{tabular}
\end{center}
\caption[]{\textit{ Same as  Figs.~\ref{fig:exclusion} but assuming that the nature's choice for the neutrino mass spectrum is the inverted hierarchy.}}
\label{fig:inverted}
\end{figure}

If the nature's choice for the neutrino mass ordering is the inverted hierarchy, the  sensitivity curves depicted in  Fig.~\ref{fig:exclusion} (a) will be shifted but in the opposite direction, see Fig.~\ref{fig:inverted} for the result from the combination of data from the Phase I of both experiments.
Notice that the method presented here is independent on the octant in which $\theta_{23}$ lies (if $\theta_{23}$ is not maximal). There exists though a minor impact on the sensitivity curves since $\theta_{23}>\pi/4$ ($<\pi/4$) implies more (less) statistics, see the first term in Eq.~(\ref{eq:probappr}).

We have explored the impact of an overall $4\%$ systematic error. The sensitivity curves, while shifted slightly for the combination of \nova at 14 km off-axis and T2K at $2.5^\circ$ off-axis (or \nova at 12 km and T2K at $2.0^\circ$), get much worst if both experiments run in their default configurations. One can understand this by making use of the bi--event neutrino--neutrino plots, see Figs.~\ref{fig:comp}: a larger error bar has an enormous impact if the ellipses width is not negligible, see Figs.~\ref{fig:comp}~(a). 

\begin{figure}[h]
\begin{center}
\begin{tabular}{ll}
\includegraphics[width=3in]{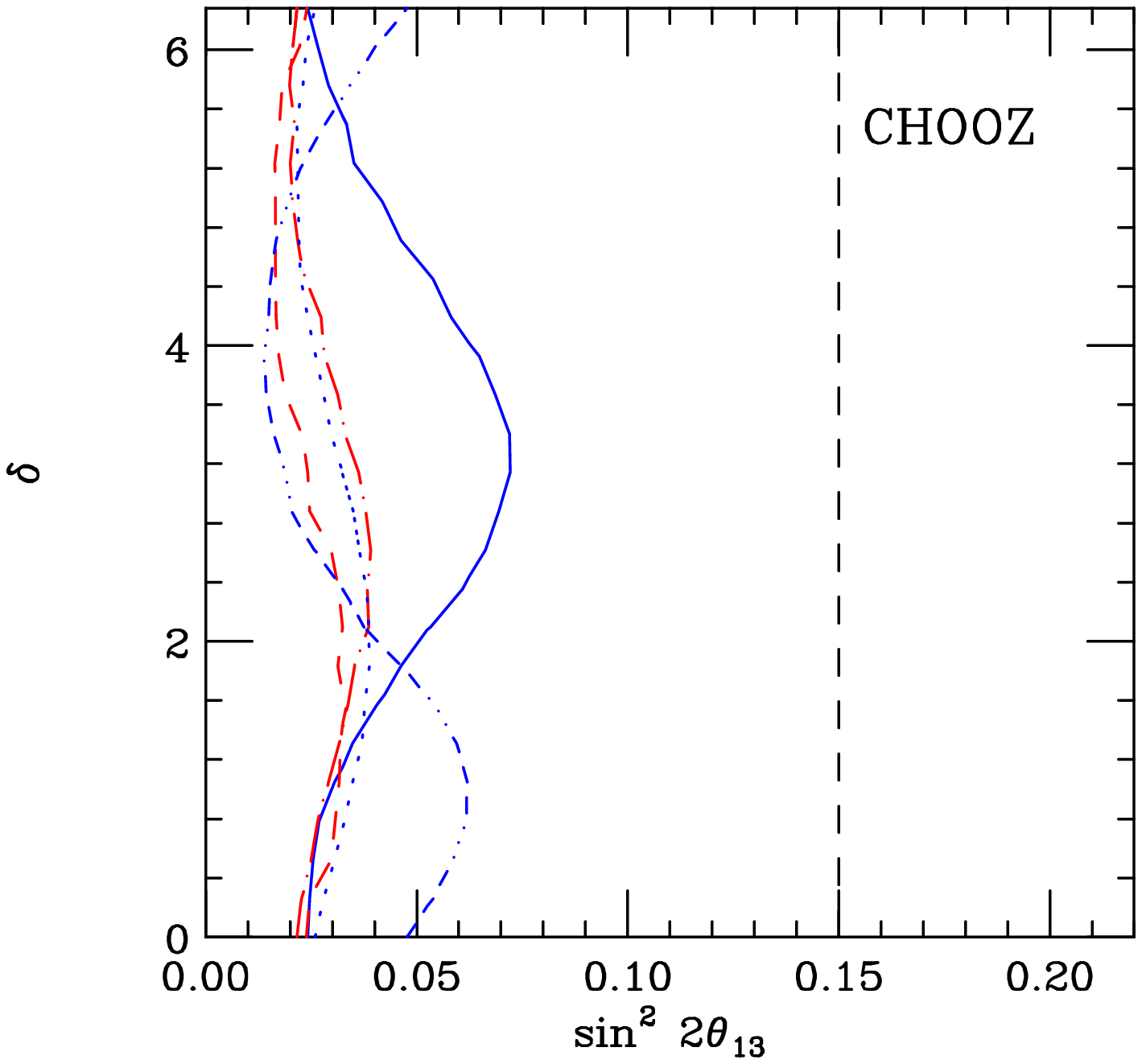}&\hskip 0.cm
\includegraphics[width=3in]{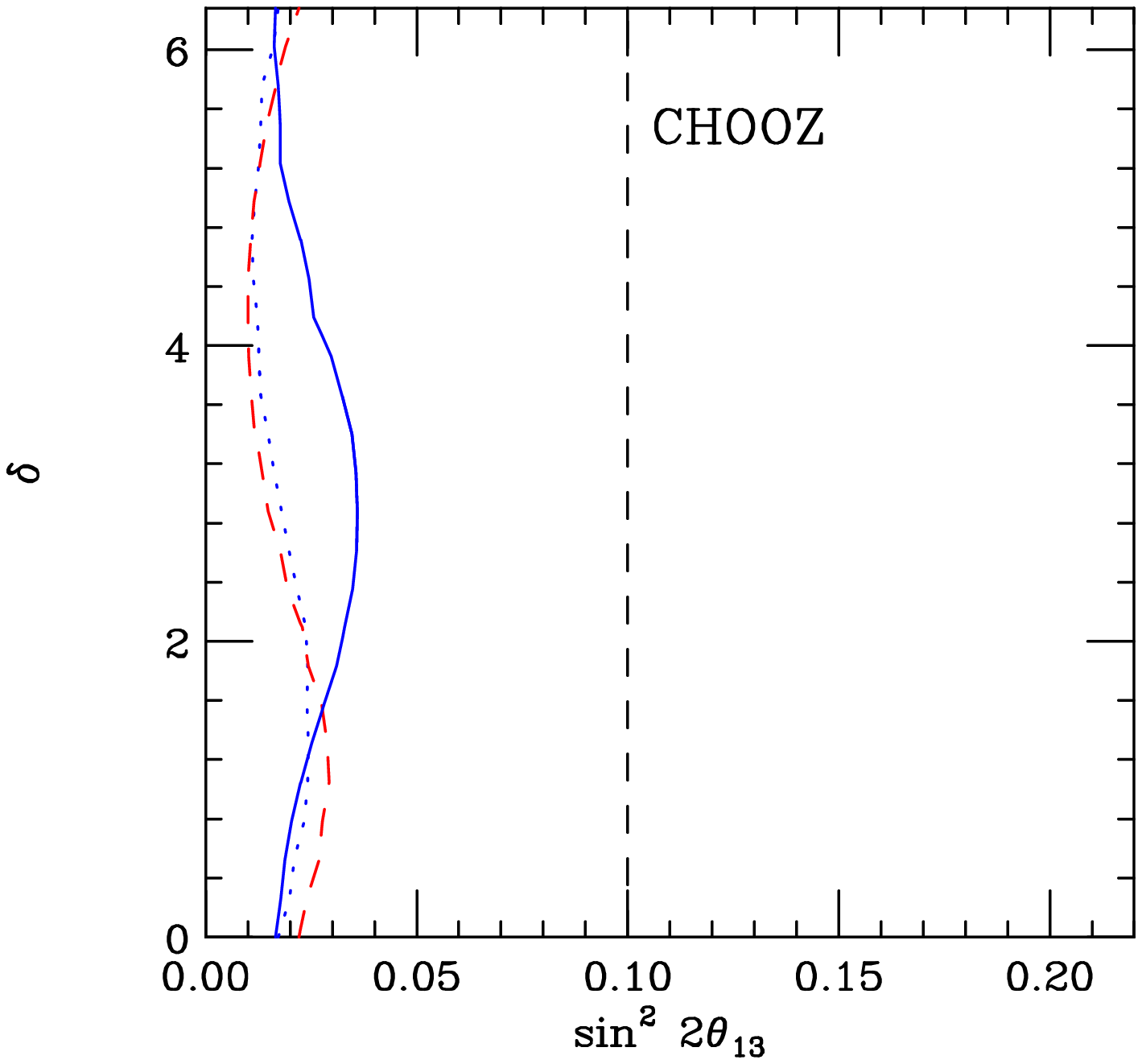}\\
\hskip 2.truecm
{\small (a) $\Delta m^2_{31} = +2.4 \times 10^{-3}$~eV$^2$}            &
\hskip 2.truecm
{\small (b) $\Delta m^2_{31} = +3.0 \times 10^{-3}$~eV$^2$}       \\  
\end{tabular}
\end{center}
\caption[]{\textit{ Same as  Figs.~\ref{fig:exclusion} but increasing the statistics of T2K and \nova by a factor of five.}}
\label{fig:exclusionbis}
\end{figure}

\begin{figure}[h]
\begin{center}
\begin{tabular}{ll}
\includegraphics[width=3in]{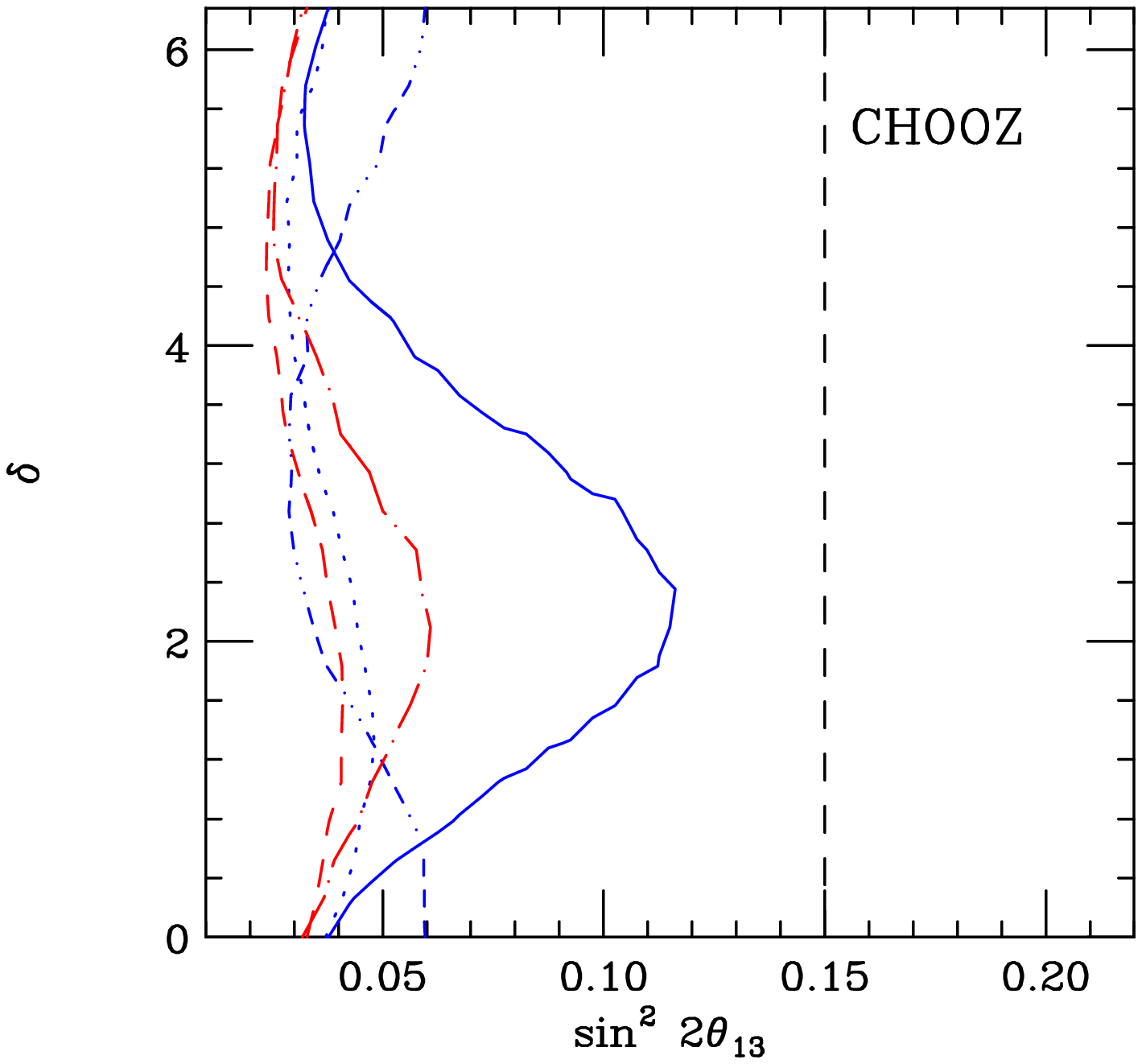}&\hskip 0.cm
\includegraphics[width=3in]{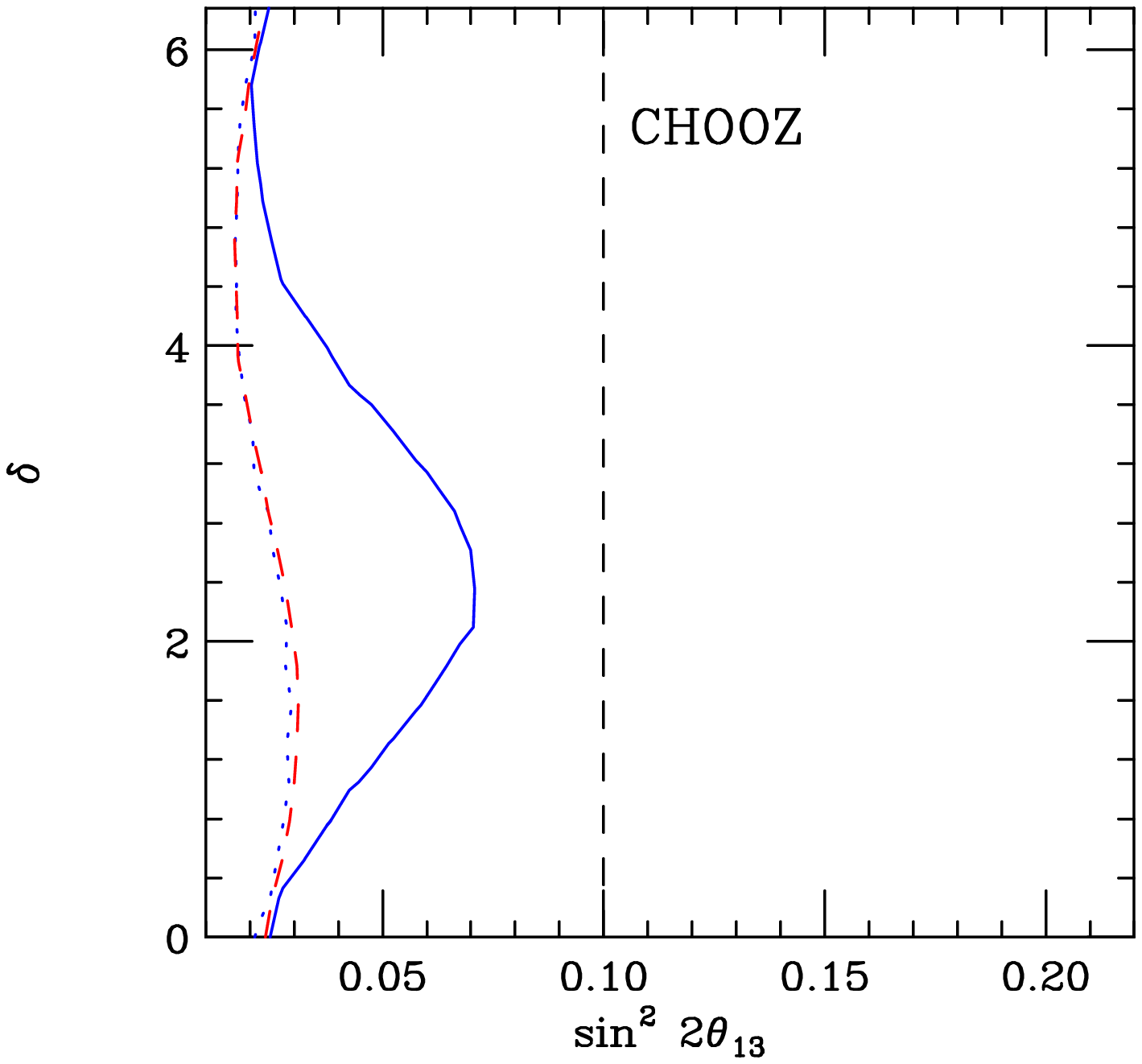}\\
\hskip 2.truecm
{\small (a) $\Delta m^2_{31} = -2.4 \times 10^{-3}$~eV$^2$}            &
\hskip 2.truecm
{\small (b) $\Delta m^2_{31} = -3.0 \times 10^{-3}$~eV$^2$}       \\  
\end{tabular}
\end{center}
\caption[]{\textit{ Same as  Figs.~\ref{fig:exclusionbis} but  assuming that the nature's choice for the neutrino mass spectrum is the inverted hierarchy.}}
\label{fig:invertedii}
\end{figure}

The combination of data from an upgraded phase of the T2K and/or \nova experiments (by increasing the proton luminosities, the years of neutrino running and/or the mass of the far detectors) will obviously increase the statistics and will shift the sensitivity curves depicted in  Fig.~\ref{fig:exclusion} (a), similarly to the effect of increasing $\Delta m^2_{31}$, see Figs.~\ref{fig:exclusionbis}, where we have upgraded \nova and T2K experiments by increasing a factor of five their expected Phase I statistics (Phase II). Fig.~\ref{fig:invertedii} depicts the results from an upgraded Phase II of both experiments in the inverted hierarchy nature's choice: if the neutrino mass hierarchy is inverted, the case for the Phase II of both experiments will be stronger, especially for $\Delta m^2_{31}=2.4 \times 10^{-3}$~eV$^2$.

\section{Conclusions}
\label{conclusions}

The most promising way to extract the neutrino mass hierarchy  is to make use of the matter effects in neutrino oscillations. For that purpose, the fastest way would be to exploit the neutrino data only from two near-term long baseline $\nu_e$ appearance experiments performed at the same $\langle E\rangle/L$, provided $\sin ^2 2 \theta_{13}$ is within their sensitivity range or within the sensitivity range of the next-generation $\bar{\nu}_e$ disappearance reactor neutrino experiments. Such a possibility could be provided by the combination of the data from the Phase I of the T2K and \nova experiments. We conclude that the optimal configuration for these experiments would be $14$~km off-axis for the \nova far detector and $2.5^\circ$ off-axis for the T2K experiment. The combination of their expected results could provide a $90\%$ confidence level (using 2 d.o.f) resolution of the neutrino mass hierarchy if $\sin^2 2\theta_{13} > 0.11$ (for $\Delta m^2_{31}=2.4\times 10^{-3}$~eV$^2$) or if $\sin^2 2\theta_{13} > 0.07$ (for $\Delta m^2_{31}=3.0 \times 10^{-3}$~eV$^2$).
A modest upgraded next Phase of both \nova and T2K experiments (by increasing a factor of five their expected Phase I statistics) could shift the $90\%$ CL limits quoted above to $\sin^2 2\theta_{13} > 0.03$ (for $\Delta m^2_{31}=2.4\times 10^{-3}$~eV$^2$) and to $\sin^2 2\theta_{13} > 0.025$ (for $\Delta m^2_{31}=3.0 \times 10^{-3}$~eV$^2$). A slightly less sensitive combination is T2K at $2^\circ$ off-axis angle and \nova at $12$ km off-axis location.  

\section*{Acknowledgements}
We acknowledge extensive discussions with Hisakazu Minakata. The calculations presented here made extensive use of the Fermilab-General-Purpose Computing Farms~\cite{farms}. Fermilab is operated by URA under DOE contract DE-AC02-76CH03000. H.~N. is supported by Funda\c{c}\~ao de Amparo \`a Pesquisa do Estado de Rio de Janeiro (FAPERJ) and Conselho Nacional  de Ci\^encia e Tecnologia (CNPq). 

\end{document}